\documentstyle[twocolumn,epsfig,aps]{revtex}
\begin{document}
 \draft
\title{A conditional quantum phase gate between two 3-state atoms}
\author{X. X. Yi, X. H. Su, and L. You}
\address{School of Physics, Georgia Institute of Technology, Atlanta GA
30332, USA}
\date{\today}
\maketitle
\begin{abstract}
We propose a scheme for conditional quantum logic between
two 3-state atoms that share a quantum
data-bus such as a single mode optical field in
cavity QED systems, or a collective vibrational state of trapped ions.
Making use of quantum interference, our scheme achieves
successful conditional phase evolution without any
real transitions of atomic internal states or populating
the quantum data-bus. In addition, it only requires
common addressing of the two atoms by external laser fields.
\end{abstract}


\narrowtext
Following the recent discovery of powerful
applications for quantum computing algorithms \cite{qa1,qa2},
quantum information science \cite{qc1} has witnessed
significant progress and development on the experimental
side. Among the variety of physical systems
being explored for hardware implementations for
quantum logic, atomic ion-traps and cavity
QED systems are favored because of their demonstrated
advantage when subjected to coherent manipulations.

Much experimental progress has been made with
trapped ions, ranging from the deterministic
creation of 4-ion internal state entanglement \cite{4ion};
the generation of arbitrary superposition states
of the quantized ion motion \cite{arbs}; to the
high fidelity quantum gate operations between
two qubits \cite{2bit}.
While a high Q optical cavity with atomic qubits inside
is an early candidate for quantum computation \cite{qed1},
it has been challenging to experimentally
realize its full promise.
In a recent paper \cite{you}, we delineated some of the
technical difficulties and proposed an easier
theoretical protocol for conditional phase dynamics
between two 4-state atoms \cite{you,dfs1}.
Let us recount the current difficulties
with cavity QED base systems;
1) precisely localizing each atomic motional wave packet
(to ensure the difficult Lamb-Dicke limit);
2) obtaining a double $\Lambda$-type, 6-state,
level diagram for each atom as required in
the protocol of Ref. \cite{qed1};
and 3) individually addressing each atom during
the gate operation when both atoms are inside the cavity.

Recent developments in
the synthesis of ion trap and cavity QED systems
have allowed attainment of the Lamb-Dicke limit \cite{tc1,tc2}.
Such a composite system suggests possibilities
for quantum communication of information stored in material
atoms through cavity photons, in addition to
processing quantum gates between several
nearby atomic qubits.

In this paper, we present a new theoretical scheme
for a quantum phase gate between two 3-state atoms.
By making use of two-atom quantum interferences,
our new protocol significantly reduces
the effect of cavity decay and the atomic spontaneous
emission during the gate operation.
Furthermore, the external control of the system is
made easier as the successful implementation of
the phase gate involves no real atomic transitions
or the presence of a cavity photon.

Our system consists of two 3-state atoms as in Ref. \cite{jane},
each with two stable ground states ($|0\rangle$ and $|1\rangle$)
for storage of one qubit of quantum information, and
an excited state ($|e\rangle$) used for virtual transitions.
In an experimental implementation,
the $M_F=0$ clock transition states of alkali metal atoms
are ideally candidates for the ground states because
of their insensitivity to residual magnetic fields.
Both atoms are assumed to couple (off-resonantly)
to an external laser field and the same single photon
mode of the cavity as illustrated in Fig. \ref{fig1}
below. The conditional phase ($\phi$) gate
then is simply given by the following unitary evolution
\begin{eqnarray}
|0\rangle_A |0\rangle_B &\to & |0\rangle_A |0\rangle_B \nonumber\\
|0\rangle_A |1\rangle_B &\to & |0\rangle_A |1\rangle_B \nonumber\\
|1\rangle_A |0\rangle_B &\to & |1\rangle_A |0\rangle_B \nonumber\\
|1\rangle_A |1\rangle_B &\to & e^{i\phi}|1\rangle_A |1\rangle_B,
\label{pg}
\end{eqnarray}
between the two atoms A and B.

\begin{figure}
{\hskip 64pt \includegraphics[width=0.85in]{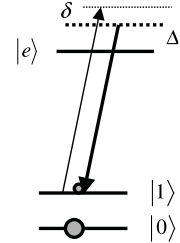}\\}
\caption{A 3-state atom interacting with
an external laser field and the cavity quantum field.}
\label{fig1}
\end{figure}

The effective model Hamiltonian of our system,
including atomic spontaneous emission and cavity decays,
is identical to that considered in Ref. \cite{jane}
for a dynamical Stark phase gate between two trapped
ions in a optical cavity.
\begin{eqnarray}
H&=&H_A+H_B+H_C,\nonumber\\
H_{\mu=A/B}&=& \hbar(\omega_e-i\frac{\Gamma}{2})
|e\rangle_\mu\langle e|+\hbar\omega_1|1\rangle_\mu\langle 1|\nonumber\\
&+&{1\over 2}[\Omega_\mu|e\rangle_\mu\langle 1|e^{-i\omega_L t}
+g_\mu|e\rangle_\mu\langle 1|c+h.c.],\nonumber\\
H_C &=& \hbar(\omega_C-i\kappa) c^{\dagger}c,
\label{ham}
\end{eqnarray}
where $H_{A/B}$ is the Hamiltonian for atom
(A/B) interacting with the cavity mode and
the external laser fields.
$\Omega_L$ is the Rabi frequency of the external laser
field ($|e\rangle\leftrightarrow|1\rangle$) at frequency
$\omega_L$ and $g_\mu$ is the single photon coherent
coupling rate ($|e\rangle\leftrightarrow|1\rangle$) of the
cavity field. $\Gamma$ denotes the atomic spontaneous emission rate
and $\kappa$ stands for the (one-side) cavity decay rate. We will not
consider the position dependance of the cavity-atom coupling
$g_\mu(\vec r_\mu)$,
a good approximation in the Lamb-Dicke limit.

Our scheme works in the following limits: 1) both
the external laser and the cavity field are strongly detuned, i.e.,
$|\Delta=\omega_C-(\omega_e-\omega_1)|\gg
\Gamma, \kappa, |\Omega_\mu|, |g_\mu|$,
and $|\delta|\sim 0$;
and 2) $|g_\mu|>|\Omega_\mu|$ and $|g_\mu|^2\gg \Gamma\kappa$
as required for the strong coupling. The dynamic Stark
gate of Ref. \cite{jane} works when the classical field is resonant
with the unperturbed atomic transition $|e\rangle\leftrightarrow|1\rangle$,
i.e. $\omega_L\sim (\omega_e-\omega_1)$, or
$\delta=\omega_L-\omega_C=-\Delta$. Therefore, it is more sensitive
to the atomic decay $\Gamma$. Our protocol, to be discussed below,
works when both fields are strongly detuned and are resonant with
each other, i.e. $\delta\sim 0$. It seems to work with
reasonable losses due to both $\Gamma$ and $\kappa$.

In the following, we analyze in detail
the two quantum interference transitions of
the combined atom + cavity system that lead to
the required conditional phase dynamics Eq. (\ref{pg}).
We assume that the atom-laser and atom-cavity
coupling strengths are adjustable so that
 $\Omega_A=-\Omega_B=\Omega$ and $g_\mu=g$.
The latter condition can be easily established if
we have the control over atomic positions inside the
cavity, e.g. a linear atomic
array intersecting the cavity at a right angle
allows the placement of the two respective atoms
in radially symmetric points of
the (cylindrically symmetric) cavity. The former condition
can be met when the plane wave laser field intersects
from the side of the cavity such that the
wave front delay between the two atoms (a distance $\vec d$ apart)
corresponds to an odd number of $\pi$, i.e.
$e^{i\vec k_L\cdot \vec d}=-1$.

\begin{figure}
{\hskip 24pt \includegraphics[width=2.in]{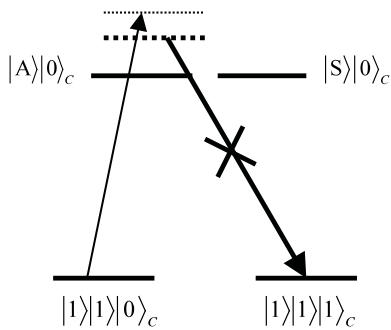}\\}
\caption{The dominant transition path starting from
state $|1\rangle_A|1\rangle_B|0\rangle_C$.}
\label{fig2}
\end{figure}

Within the above specific regime of parameters, the
state $|0\rangle_A|0\rangle_B$ experiences no dynamics
because the hyperfine splitting $\omega_{10}$ is much
larger than $|\Delta|$ in all practical implementations.
The state $|1\rangle_A|1\rangle_B$, on the other
hand, accumulates a phase shift that increases linearly with time.
As illustrated in Fig. \ref{fig2}, it couples
first to the anti-symmetric one atom excited state
$|{\cal A}\rangle=(|e\rangle_A|1\rangle_B-|1\rangle_A|e\rangle_B)/\sqrt{2}$
because the choice of $\Omega_\mu$. The state $|{\cal A}\rangle$,
however, CANNOT couple to the one cavity photon state
$|1\rangle_A|1\rangle_B|1\rangle_C$ because of the choice
$g_\mu=g$. In fact, state $|1\rangle_A|1\rangle_B|1\rangle_C$
is only coupled to the symmetric state
$|{\cal S}\rangle=(|e\rangle_A|1\rangle_B+|1\rangle_A|e\rangle_B)/\sqrt{2}$.
Further coupling to the two-atom excited state
from $|{\cal A}\rangle$ is weakened because of the
twice as large detuning.
Thus the only dynamics for state $|1\rangle_A|1\rangle_B$
is a constant rate of its phase due to the
AC Stark shift induced by the classical field $\Omega$,
which we find to be
$\phi(t)=-{|\Omega|^2\over 2(\Delta+\delta)}t$ when both
decay rates are ignored.

\begin{figure}
{\hskip 12pt \includegraphics[width=2.25in]{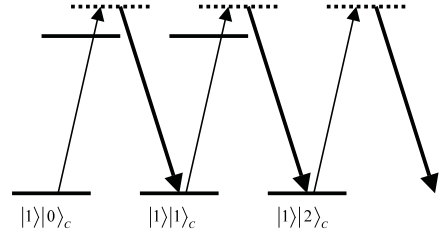}\\}
\caption{The dominant single atom transition paths
starting from atomic states $|1\rangle_A|0\rangle_B$
or $|0\rangle_A|1\rangle_B$ and an empty cavity.}
\label{fig3}
\end{figure}

Now let's look at states
$|0\rangle_A|1\rangle_B$ and $|1\rangle_A|0\rangle_B$.
It turns out that they accumulate no phase shifts
at all as long as $\delta\sim 0$, again due to
an interesting quantum interference.
In this case the dominant transition paths are
the single photon resonances between the classical
and cavity fields as illustrated in Fig. \ref{fig3}.
In the limit of $|g|\gg |\Omega|$, these resonant
transitions become
essentially summed by adopting the
dressed state basis as in the standard Jaynes-Cummings model.
All excited states then assume a doublet manifold
given by
\begin{eqnarray}
|+,n\rangle &=&
\sin\theta|1\rangle|n+1\rangle+\cos\theta|e\rangle|n\rangle,\nonumber\\
|-,n\rangle &=& \cos\theta |1\rangle|n+1\rangle-\sin\theta
|e\rangle|n\rangle,
\end{eqnarray}
with the corresponding eigenvalues
\begin{eqnarray}
E_{\pm}(n)=-\frac{\Delta}{2}\pm\sqrt{\frac{\Delta^2}{4}+|g|^2(n+1)}.
\end{eqnarray}
The mixing angle $\theta$ is determined from
$\tan (2\theta)=-{2g\sqrt{n+1}}/{\Delta}$.
For the first excited state ($n=0$),
the above eigven-values also represent the detunings
of states $|\pm,0\rangle$ from the external laser
field at $\omega_L=\omega_C$.

\begin{figure}
{\hskip 12pt \includegraphics[width=2.in]{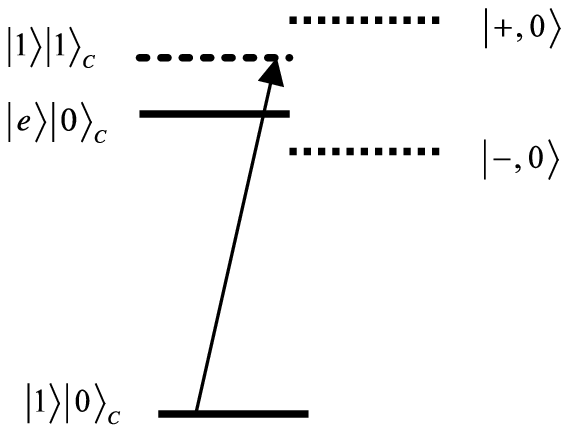}\\}
\caption{The dominant single atom transition paths
starting from atomic state $|1\rangle_A|0\rangle_B$
or $|0\rangle_A|1\rangle_B$ and an empty cavity,
in the dressed state picture.}
\label{fig4}
\end{figure}

The weak classical field $\Omega$ then simply drives
the atom from the ground state $|1\rangle_\mu|0\rangle_C$
to the first excited state doublet as in Fig. \ref{fig4}.
Higher excited doublets are ignored
again because of the large detuning.
The AC Stark shift
of the ground state $|1\rangle_A|0\rangle_C$
then simply results from the summation of the
respective shifts of the doublet $|\pm,0\rangle$.
It is an easy exercise to show that the shift from
the upper state $|+,0\rangle$ (negatively detuned)
exactly cancels that from the lower state
$|-,0\rangle$ (positively detuned), i.e.
\begin{eqnarray}
\delta_{10}=-\frac{|\Omega|^2 \cos^2\theta}{4E_+(0)}
-\frac{|\Omega|^2 \sin^2\theta}{4E_-(0)}=0.
\end{eqnarray}

We have performed extensive numerical simulations
with the full Hamiltonian Eq. (\ref{ham}).
When both decays are ignored by
taking $\Gamma=\kappa=0$, we find the above analytical
insights to be completely accurate, i.e. we indeed
execute the conditional phase gate Eq. (\ref{pg}).
In fact, we find that the gate works with almost
perfect ($>99\%$ fidelity) even beyond the limit
our interference based analysis implies,
except that states $|1\rangle_A|0\rangle_B$
and $|0\rangle_A|1\rangle_B$ now also accumulate phases
$\phi_A(t)$ and $\phi_B(t)$ respectively.
The conditional phase then becomes $\phi(t)-\phi_A(t)-\phi_B(t)$.
The phase gate of Ref. \cite{jane}
involves a Rabi oscillation between state
$|1\rangle_A|1\rangle_B$ and $|{\cal A}\rangle$.
Thus it is less tolerant to residual phase shifts
for states $|1\rangle_A|0\rangle_B$
or $|0\rangle_A|1\rangle_B$.

\begin{figure}
{\hskip 12pt \includegraphics[width=3.25in]{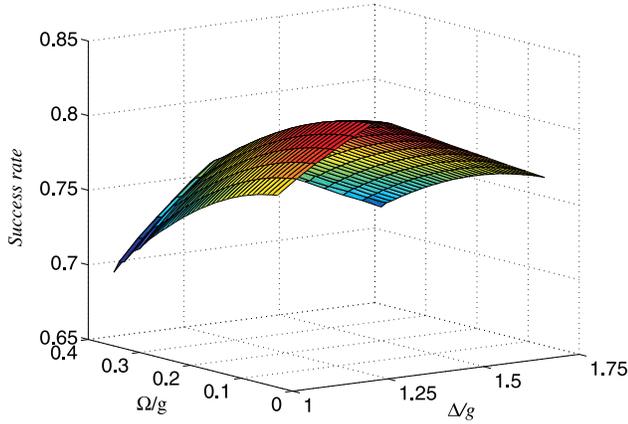}\\}
\caption{The success rate for selected system parameters.}
\label{tfig1}
\end{figure}

Figure \ref{tfig1} shows selected results for the dependence of
the phase gate success rate on system parameters for
$\Gamma=0.08|g|$ and $\kappa=0.05|g|$ ($|g|^2=250\Gamma\kappa$).
Quite promising results are obtained over a broad range of
system parameters. The correspondingly fidelities are always
approaching being perfect (${\cal F}\ge 99.9\%$).

\begin{figure}
{\includegraphics[width=3.in]{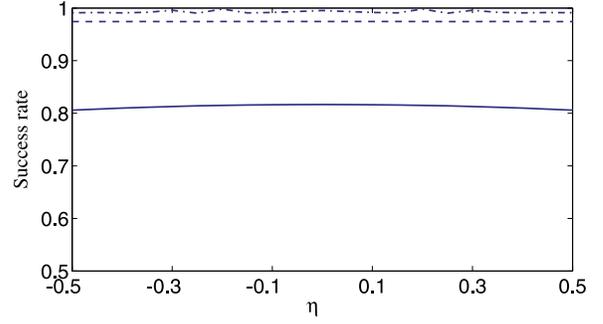}\\}
\caption{The dependence of optimal
success rate on $\eta$
for several choices of system parameters:
$\Gamma=\kappa=0$ (dot-dashed),
$\Gamma=0.05|g|$, $\kappa=0$ (dashed line), and
$\Gamma=0.08|g|$, $\kappa=0.05|g|$ (solid line).}
\label{tfig2}
\end{figure}

In Fig. \ref{tfig2}, we display the dependence of
the optimal success rate on the relative phase $\eta$
between $\Omega_A=e^{i\eta}\Omega$ and $\Omega_B=-\Omega $.
We see that the dependence is essentially flat.
For $\eta\neq 0$, transitions from
state $|1\rangle_A|1\rangle_B$ in Fig. \ref{fig2}
become modified in similar ways to those in Fig. \ref{fig3}
except the excited state is now a linear combination
of $|{\cal A}\rangle$ and $|{\cal S}\rangle$. A similar
reduction to doublet structures as in Fig. \ref{fig4}
occurs since $|g|\gg |\Omega_\mu|$. Thus the complete
diagram is still essentially closed; now, however, the
phase accumulation rate becomes smaller due to
the cancellation of AC Stark shifts from the
first excited doublet.

\begin{figure}
{\hskip 12pt \includegraphics[width=2.in]{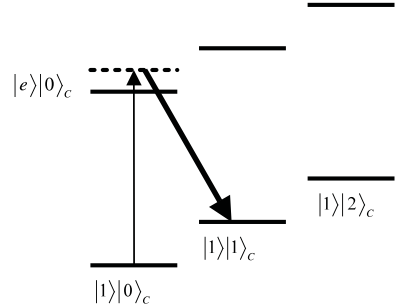}\\}
\caption{An equivalent configuration of Fig. \ref{fig1}
for trapped ions. The internal state $|0\rangle$ and its associated
vibrational band is not drawn as $\omega_{10}\gg \omega_C$.
}
\label{fig5}
\end{figure}

Although the limit of $|g|^2\gg\Gamma\kappa$ is a
challenging pursuit \cite{tc5}, it can nevertheless
be expected to be reachable
with optical cavity QED based systems soon \cite{tc3,tc4}.
On the other hand, for the state of art ion traps,
$\kappa$ can be made exceedingly small and
$\Gamma$ can be lowered using Raman transitions \cite{4ion}.
We now show that
our model maps perfectly to an ion trap setup \cite{ther3}.
For the same type 3-state atoms as illustrated in Fig. \ref{fig1},
the sideband of their collective vibrational mode
is shared by the two atomic ions.
We will use two external classical fields;
one running wave drives the single atomic ion
(off-resonantly) at the carrier
frequency $|1\rangle_{\mu}|n\rangle_C\leftrightarrow
|e\rangle_{\mu}|n\rangle$ (insensitive to the motion);
and the other drives the red motional sideband as
illustrated in Fig. \ref{fig5}.
The equivalent
cavity coupling $g_\mu$ then becomes in this case
$R_\mu$, the Rabi frequency for the second
laser (much stronger than the first one at the carrier frequency)
times its respective Lamb-Dicke parameters of
the harmonic vibration for each of the atoms $\eta_\mu$ \cite{ab}.
The Hamiltonian of the system is then again described
by Eq. (\ref{ham}), except now the index `C' refers to
the collective vibrational state and $c^\dag$ ($c$)
denotes the creation (annihilation) of a
vibration quanta. We will require
$|\Delta|\ll \omega_C$ to prevent any internal excited
state with a non-zero vibrational quanta from participating
the gate dynamics.

If the collective center of mass vibrational mode is used,
then $\eta_\mu=\eta$, the same as in the cavity QED model.
We could also use the two ion breathing mode, then
$\eta_A=-\eta_B$, which leads to $g_A=-g_B=g$.
If we then take $\Omega_A=\Omega_B=\Omega$ instead,
the same theory as developed above applies.

\begin{figure}
{\includegraphics[width=3.in]{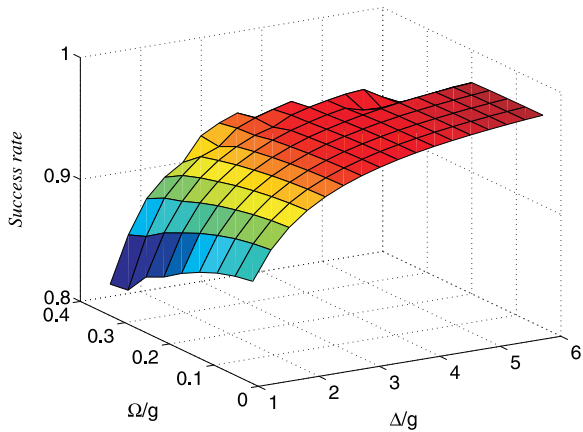}\\}
\caption{Same as in Fig. \ref{tfig1} but for
$\Gamma=0.05|g|$ and $\kappa=0$.}
\label{tfig3}
\end{figure}

In Fig. \ref{tfig3}, we show selected results
of the success rate for $\kappa=0$, a situation
close to an ion trap implementation.

In conclusion, we have discovered a new
protocol for performing a quantum phase gate
between two atomic qubits. We have explained
its operating mechanism in terms of
quantum interference effects. In addition,
our protocol can also be explained in terms of a
quantum Zeno subspace \cite{qzs1} or an environment
induced decoherence free space as in other cavity QED
based theories \cite{jane,other}. We
believe this new protocol is more advantageous
when compared to other known cavity QED protocols.
It requires only common addressing of the two atoms
during the gate operation;
it uses 3 internal states for each atomic qubit,
and applies to both cavity QED and trapped ion systems;
Furthermore, its successful implementation
involves no real transitions of atomic states
or populating the cavity or the collective motional
state quantum data-bus. In the ideal
operation limit, this protocol thus becomes
reasonably insensitive to motional
dephasing or heating because the Stark phase
due to an external plane wave laser is position independent.

We thank Dr. M. Plenio for helpful communications.
This work is supported by a grant from NSA, ARDA, and DARPA
under ARO Contract No. DAAD19-01-1-0667, and by a grant
from the NSF PHY-0113831.


\begin{thebibliography}{99}
\bibitem{qa1}P. W. Shor, Proc. 35th Ann. Symp. on the Foundations
of computer science, Ed. S. Goldwasser, (Los Alamitos, CA IEEE
computer society).

\bibitem{qa2}L. K. Grover, Phys. Rev. Lett. {\bf 79}, 325(1997).

\bibitem{qc1}D. Deutsch, Proc. R. Soc. A {\bf 400}, 97(1985);
D. Deutsch, Proc. R. Soc. A {\bf 425}, 73(1989).

\bibitem{4ion}C. A. Sackett {\it et al.},
Nature(London) {\bf 404}, 256(2000).

\bibitem{arbs}
A. Ben-Kish, B. DeMarco, V. Meyer, M. Rowe, J. Britton,
W.M. Itano, B.M. Jelenkovi\'c, C. Langer, D. Leibfried,
T. Rosenband, and D.J. Wineland, (quant-ph/0208181).

\bibitem{2bit}
B. DeMarco, A. Ben-Kish, D. Leibfried, V. Meyer, M. Rowe, B.M. Jelenkovi\'c,
W.M. Itano, J. Britton, C. Langer, T. Rosenband, and D.J. Wineland,
(quant-ph/0208180).

\bibitem{qed1}T. Pellizzari, S. Gardiner, J. I. Cirac, and P. Zoller,
Phys. Rev. Lett. {\bf 75}, 3788(1995).

\bibitem{you}L. You, X. X. Yi, and X. H. Su,
(quant-ph/0209096).

\bibitem{dfs1}J. Pachos and H. Walther, (quant-ph/0111088).

\bibitem{tc1}G. R. Guthohrlein, M. Keller, K. Hayasaka, W. Lange,
and H. Walther, Nature {\bf 414}, 49(2001).

\bibitem{tc2}A. B. Mundt, A. Kreuter, C. Becher, D. Leibfried,
J. Eschner, F. Schmidt-Kaler, and R. Blatt,
Phys. Rev. Lett. {\bf 89}, 103001 (2002).

\bibitem{jane}E. Jane, M. B. Plenio, and D. Jonathan,
Phys. Rev. A {\bf 65}, 050302 (2002).

\bibitem{tc5}
P. Bertet, S. Osnaghi, P. Milman, A. Auffeves, P. Maioli, M.
Brune, J. M. Raimond, and S. Haroche, Phys. Rev. Lett. {\bf 88},
143601 (2002).

\bibitem{tc3}C. J. Hood {\it et al}., Science {\bf 287}, 1447 (2000);
A. C. Doherty, T. W. Lynn, C. J. Hood, and H. J. Kimble, Phys.
Rev. A {\bf 63}, 013401 (2001).

\bibitem{tc4}P. W. H. Pinkse {\it et
al.}, Nature {\bf 404}, 365 (2000); M. Hennrich
{\it et al.}, Phys. Rev. Lett. {\bf 85}, 4872 (2000).

\bibitem{ther3}S. Schneider, D. F. V. James, and G. J. Millburn,
J. Mod. Opt. {\bf 47}, 499 (2000).

\bibitem{ab}Almut Beige, (quant-ph/0205070).

\bibitem{qzs1} P. Facchi and S. Pascazio,
Phys. Rev. Lett. {\bf 89}, 080401 (2002).

\bibitem{other}
A. Beige, D. Braun, B. Tregenna, and P. L. Knight,
Phys. Rev. Lett. {\bf 85}, 1762 (2000).

\end{thebibliography}
\end{document}